\documentclass [pra, twocolumn, aps, showpacs
] {revtex4}
\usepackage{graphicx}
\usepackage{amsmath}
\usepackage{amssymb}

\def\gsim{\mathrel{\rlap{\lower4pt\hbox{\hskip1pt$\sim$}}\raise1pt\hbox{$>$}}}

\begin{document}

\title{Evolution of the Pseudogap in a polarized Fermi gas}

\author{Erich J. Mueller}
\affiliation{Laboratory for Atomic and Solid State Physics, Cornell
University, Ithaca NY 14853}

\date{Dec 3, 2010}

\begin{abstract}
We calculate the single particle spectral density of a normal (non-superfluid) two component gas of fermions in the BCS-BEC crossover within a T-matrix approximation.  We review how non-condensed pairs lead to a spectral density reminiscent of the ordered state, and explore how a gap-like feature in the spectrum evolves as one changes the polarization of the gas.  As the gas is polarized we find that this pseudogap becomes more diffuse and moves away from the Fermi level, reflecting the fact that fewer pairs are present but that they still play an important role in the excitations.  
\end{abstract}


\maketitle

Two closely related themes in modern condensed matter physics are how correlations manifest themselves in unordered states of matter, and the existence of materials with unordered ground states.  These themes are ubiquitous, playing a role in studies of spin systems, transition metal oxides, and cold gases: they are key to phenomena such as spin liquids and resonating valence bonds \cite{wenreview}.  One important idea to emerge in this context is that of a ``pseudogap": the fact that under appropriate circumstances the normal state displays a suppression of the single particle spectral density near the Fermi level, reminiscent of the gaps seen in ordered states of matter.  Here we present a theoretical study of  the pseudogap in a strongly interacting Fermi gas, such as $^6$Li or $^{40}$K atoms.
In particular, we show how this pseudogap structure evolves as one polarizes the gas.

 Over the past decade, the pseudogap system which has received the most attention has been high temperature cuprate superconductors \cite{cuprates}.  In that system the origin of the pseudogap is widely debated, with some arguing that it is actually a real gap associated with non-superfluid ordering.  Less controversially, there are at least two other systems where 
 vestiges of  ordering dominate the properties of the normal state:
 atoms in the ``BCS-BEC" crossover \cite{latestjin,chienlevin,hu,ohashi,strinati,levincompare,haussmann,stajic,bcsbecpseudo,perali,janko,trivedi,attractivehubbard,sheehy,levinrf,bcsbecrf,levinvarena}, and electrons in one-dimensional charge-density wave materials \cite{cdw}.  The former system is ideal for studying pseudogap phenomena:  one has a wealth of ``knobs" for adjusting the parameters, and a number of very powerful probes.  For example, we will calculate the single particle spectral density, which can be extracted from RF spectroscopy experiments \cite{levinrf,bcsbecrf,latestjin}.

 These experiments on cold fermi gases have dramatically shifted our perspective on superconductivity/superfluidity \cite{bcsbecrf,bcsbecexp}.  They have confirmed the ideas, largely developed in the 1980's, that the superfluidity seen in $^4$He -- where bosonic atoms condensed -- is continuously connected to the superfluidity seen in $^3$He -- where superfluidity results from a fermi surface instability \cite{bcsbecearly1,bcsbecearly2,nozieres}.  In the cold gas experiments a magnetic field is used to tune the energy of a two-body state \cite{feshbachreview}: taking it from a negative value (where it is a bound state) to a positive value (where it is a scattering resonance).    In the former ``BEC" limit, the low temperature system is described as a gas of bosonic dimers, while in the latter ``BCS"  limit it is a gas of fermions, interacting via an effective attractive interaction.  The superfluid ground state smoothly evolves as one changes the magnetic field:  the molecular pairs on the BEC side simply grow into large Cooper pairs as one approaches the BCS limit.  In this crossover, one of the most interesting point is where the free-space bound state has exactly zero energy.  Here interactions provide no length-scale, and the scattering cross-section saturates the bounds set by the unitarity of the S-matrix.  At this ``unitary point," the thermodynamics potentials are ``universal" functions of the dimensionless combinations $\lambda^3 n_\sigma$, where $n_\sigma$ is the density of atoms with spin $\sigma$ and $\lambda^2=2\pi\hbar^2/mk_BT$ is the thermal wavelength, where $\hbar=h/2\pi$ is the reduced Planck's constant, $m$ is the atomic mass, $k_B$ is Boltzmann's constant, and $T$ is the temperature \cite{universal}.  The experimental systems are remarkably stable near unitarity.

 One difficulty in studying the pseudogap is the absence of a single universally accepted definition.  In the context of high temperature superconductors the term is typically used to refer to
 a number of phenomena in the underdoped normal state.  Only some of these properties are found in the BCS-BEC crossover (see \cite{randeriavarena,levinvarena}  for reviews).  
 Throughout this paper I will define the pseudogap in terms of a dip in the density of states.  This tends to be one of the weaker 
 definition of the pseudogap in the BCS-BEC crossover, and the one which is common to all sufficiently sophisticated 
 theories of the normal state near $T_c$.
Other possible definitions, such as bimodality of the momentum resolved single particle spectral density at the Fermi energy \cite{levincompare,sheehy,ohashi}, or ``back-bending" of the peak in the single particle spectral density \cite{latestjin}, are more stringent: not all theories display them.  For example, the T-matrix theory used in this paper  does not show a definitive ``back-bending" structure \cite{levincompare}, especially when the gas is polarized.  In response to this plurality of definitions, Tsuchiya et al. \cite{ohashi} introduced the phenomonological concept of two ``pseudogap temperatures", one associated with the appearance of a dip in density of states, and another with the appearance of ``back-bending".

Alternatively, some use non-spectroscopic signatures to define the pseudogap.  For example, Shin
defines the pseudogap state in terms of the equation of state \cite{shin}.  According to his definition, 
 the temperature dependence of the
pressure found by
Nascimb\`ene et al \cite{nascimbene}
is incompatible with a pseudogap.  This interpretation is quite contentious: both Chien and Levin \cite{levinfl} and Perali et al. \cite{latestjin} point out that strong pairing correlations in the normal state can lead to gap-like features in the spectrum yet produce thermodynamic functions which are consistent with the experiments.

%
%

 \begin{figure}[tbp]
 \vspace{0.5cm}
 \includegraphics[width=\columnwidth]{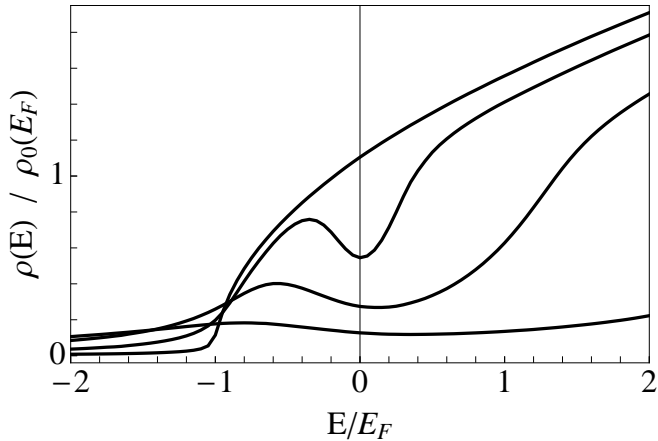}
 \caption{Pseudogap in the density of states of the (two component spin balanced) normal state just above $T_c$, calculated within our T-matrix approximation.  Vertical Axis: density of states $\rho(E)$ measured in terms of the density of states at the Fermi surface of a non-interacting gas $\rho_0(E_F)$.  From top to bottom (at $E=0$), $1/k_F a=-2.6,-0.7,-0.3,0.0$.  All energies are measured from the chemical potential. 
  The partial self-consistency enhances the suppression of spectral weight for strong interactions, broadening the pseudogap (see related figures in Refs. \cite{ohashi,bcsbecpseudo,haussmann,levincompare}, which show how varying degrees of self-consistency influence the spectra).
 }\label{dos}
 \end{figure} 

The experimental evidence for the pseudogap in Fermi gases comes from RF spectroscopy measurements \cite{latestjin}.  In those measurements one extracts the trap averaged, but momentum resolved, density of states.  At each $k$, the experimentalists find the energy, $E_k$, for which the  density of states is maximal.  In both the superfluid and normal state they find this dispersion is non-monotonic.  Fitting the dispersion to a BCS form $E_k^2=(k^2-\mu)^2+\Delta_{\rm pg}^2,$ with free parameters $\mu$ and $\Delta_{\rm pg}$, gives an estimate of the ``pseudogap" energy scale, $\Delta_{\rm pg}$.


%

 The physical picture used to understand the 
 pseudogap in the BEC-BCS crossover is that of ``preformed pairs".
 This is conceptually clearest deep on the BEC side of resonance, where the normal state is 
 %
 described as a non-condensed gas of bosonic dimers (the ``preformed pairs").  Any single-fermion excitations requires breaking a pair: either by supplying the energy with your probe, or by relying on a thermal fluctuation.  The former excitations have gaps, while the latter ones are exponentially rare for $T$ below the binding energy $E_b=k_b T^*$.  Thus there is an exponentially small density of states at the Fermi energy.  As one approaches the BCS side of resonance, the temperature scale for pairing,  $T^*$, drops towards $T_c$, and the pseudogap regime $T_c<T<T^*$ becomes vanishingly small.

 Here we ask what happens to the pseudogap when the atomic gas is ``spin imballanced" or polarized, meaning that $n_\uparrow>n_\downarrow$.  Experimentally one sees that as the polarization increases the transition temperature drops.  This trend is not surprising:  fewer $\downarrow$-particles means that it will be harder to form pairs.  When $n_\uparrow/n_\downarrow=\gamma$ exceeds a critical value $\gamma^*\approx 2.3$, the gas remains normal even at zero temperature \cite{nascimbene,critpol,parish}.

 \begin{figure}[tbp]
 \includegraphics[width=\columnwidth]{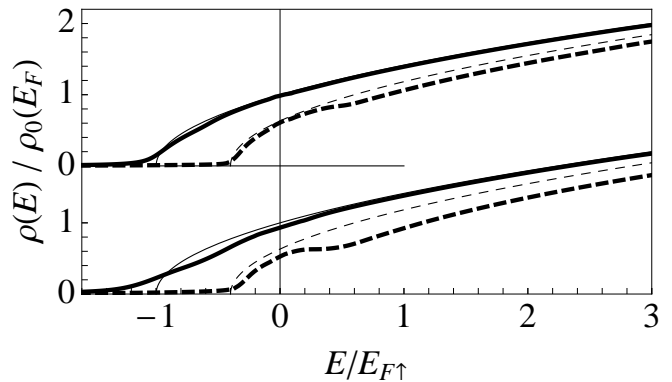}
 \caption{Density of majority species (solid) and minority species (dashed) states in the zero temperature normal state of a spin imbalanced Fermi gas.
 Top: $1/(k_{f\uparrow} a)=-0.7$, Bottom: $1/(k_{f\uparrow} a)=-0.35$;  $n_\downarrow/n_\uparrow=0.25$.  Energies are measured from the chemical potential of each species, thin lines show noninteracting density of states.}
 \label{zerotdos}
 \end{figure}
 
 Does this zero temperature normal gas have a pseudogap?  Given that one typically interprets the pseudogap in terms of pairs, the presence of a pseudogap would nominally imply that there are non-condensed pairs in this zero temperature gas.  A zero temperature non-condensed Bose gas is extremely exotic: Mott insulators, and bosonic fractional quantum Hall states, are two examples that come to mind.
 
 Since such an exotic state seem unlikely, one expects not to see a pseudogap in the polarized gas.  How does this happen?  Where does the pseudogap go as one continuously tunes the system from the high temperature unpolarized normal state to the low temperature polarized normal state?

\begin{figure*}
\vspace{-0.1in}
\includegraphics[width=0.8 \textwidth]{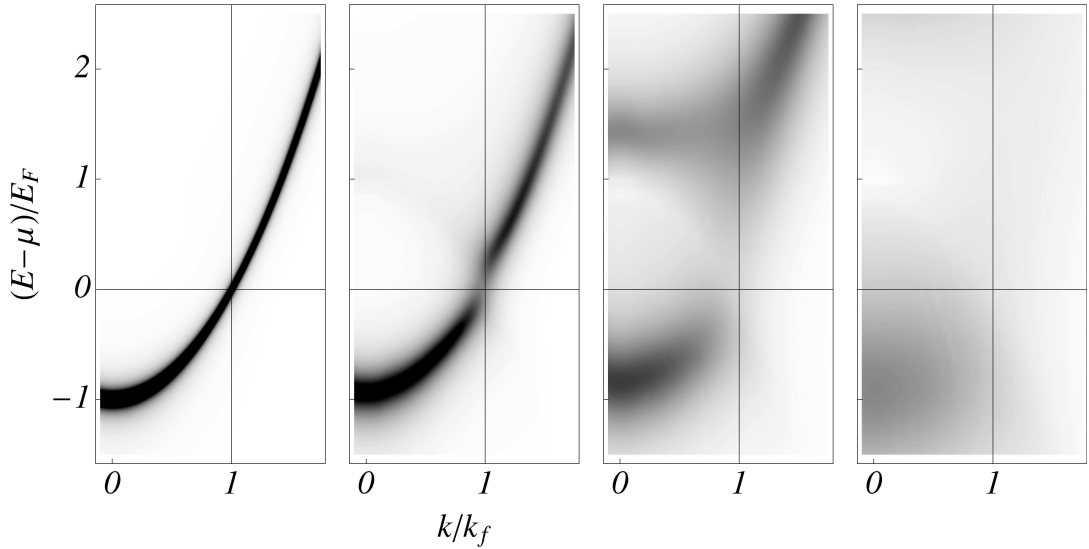}
\vspace{-0.35in}
\caption{(Wide) Single particle spectral density in the normal state of an unpolarized two-component Fermi gas with $T$ slightly above $T_c$.  Darker colors represent higher density of states.   Left to right -- weak to strong interactions $1/k_f a=-2.6,-0.7,-0.3,0.0$.   }\label{sd}
\end{figure*}

 Our answers to these questions are summarized by the density of states in Figs.~\ref{dos} and \ref{zerotdos}.  Figure~\ref{dos} illustrates that there is a significant depression in the unpolarized normal-state density of states at the Fermi energy when $T\gsim T_c$.  This depression is how we define the pseudogap.  Fig.~\ref{zerotdos} shows that in the zero-T spin-imbalanced gas this depression is pushed away from the Fermi energy.  The spectral dip is  at positive energy for the minority species density of states and negative energy for the majority species.  These very broad depressions are continuously connected to the spectral dips in Fig.~\ref{dos}.  
 The remainder of this paper describes the physics behind this finite energy pseudogap, and details our calculation of the spectral density.   We will argue that like the pseudogap in Fig.~\ref{dos}, the spectral features in Fig.~\ref{zerotdos} arise from pairing physics.  No pairs occur in the ground state, but the pairs play an important role in the excitations.

 We take a single-channel model, with Hamiltonian $H=K+V$, where $K=\sum_{k\sigma} \epsilon_{k\sigma} a_k^\dagger a_k$ and $V=g\sum_{kpq}a_{k\uparrow}^\dagger a_{p\downarrow}^\dagger a_{p-q\downarrow}a_{k+p\uparrow}$, the dispersion is $\epsilon_{k\sigma}=k^2/2m-\mu_\sigma$ and the coupling constant
$g$ is related to the scattering length by $g^{-1}=m/(4\pi\hbar^2 a_s)-\frac{1}{V}\sum_q\frac{m}{q^2} $.
 We use a T-matrix approximation to the single-particle self-energy,
 \begin{eqnarray}\label{tmateqn}
 \Sigma_\uparrow(k,\omega_n)&=&\frac{1}{\beta \Omega}\sum_{q,\nu_n} T_q(\nu_m)G^{(0)}_{q-k\,\downarrow}(\nu_m-\omega_n)
 \nonumber\\
  T^{-1}_k(\nu)&=&\frac{m}{4\pi \hbar^2 a}+\Theta_k(\nu)\\\nonumber
 \Theta_k(\omega_n)&=&
 \frac{1}{\beta \Omega}
 \sum_{q,\nu_m} 
 G^{(0)}_{k/2+q\,\uparrow}(\nu_m)
 G^{(0)}_{k/2-q\,\downarrow}(\omega_n-\nu_m)
 \end{eqnarray}
 where $\Omega$ is the volume of space, $\beta=1/k_B T$ is the inverse temperature, the frequency sums run over Matsubara frequencies, $\omega_n=2\pi i m/k_B T$.  The free Greens functions are
 \begin{equation}\label{freeg}
 G^{(0)}_{k\sigma}(\omega)=1/(\omega-\epsilon_{k\sigma}).
 \end{equation} 
 
 Physically this approximation amounts to solving the two-body problem, taking into account the presence of all of the other atoms only through Pauli blocking.
 This is the minimal model for capturing the BEC-BCS crossover: the corresponding free energy was introduced by Nozieres and Schmidt-Rink \cite{nozieres}, and it has been used/modified by numerous authors \cite{chienlevin,hu,ohashi,strinati,levincompare,haussmann,stajic,perali,janko,bcsbecpseudo,bcsbecearly1,parish,tcs}.  Mathematically it can be derived in a functional integral formulation by taking Gaussian pairing fluctuations about the noninteracting normal state.  Although one has no {\em a-priori} reason to expect that this approximation is quantitatively accurate when the interactions are strong, it appears to work surprisingly well: 
 it predicts $T_c\approx0.2E_f$ at unitarity, while
   Monte-Carlo \cite{prokof} finds $T_c=0.152(7)E_f$.  As described below, we will be using a modified version of this theory for which quantitative predictions are less accurate, but which avoids some unphysical features.  For qualitative matters, such as the one concerning us, our modified theory should  be reliable.  The theory in (\ref{tmateqn}) 
 has been used extensively to investigate pseudogap physics  in the unpolarized case \cite{ohashi}, where it displays features which are similar to the fully self-consistent T-matrix approximation \cite{haussmann}.  As discussed by Chien, Guo, He and Levin \cite{levincompare}, different levels of self-consistency yeild stronger/weaker spectral features.  The simple approach in Eq.~(\ref{tmateqn}) yields a particularly difuse pseudogap, which is further broadened by our partial self-consistency.

 \begin{figure}
\includegraphics[width=\columnwidth]{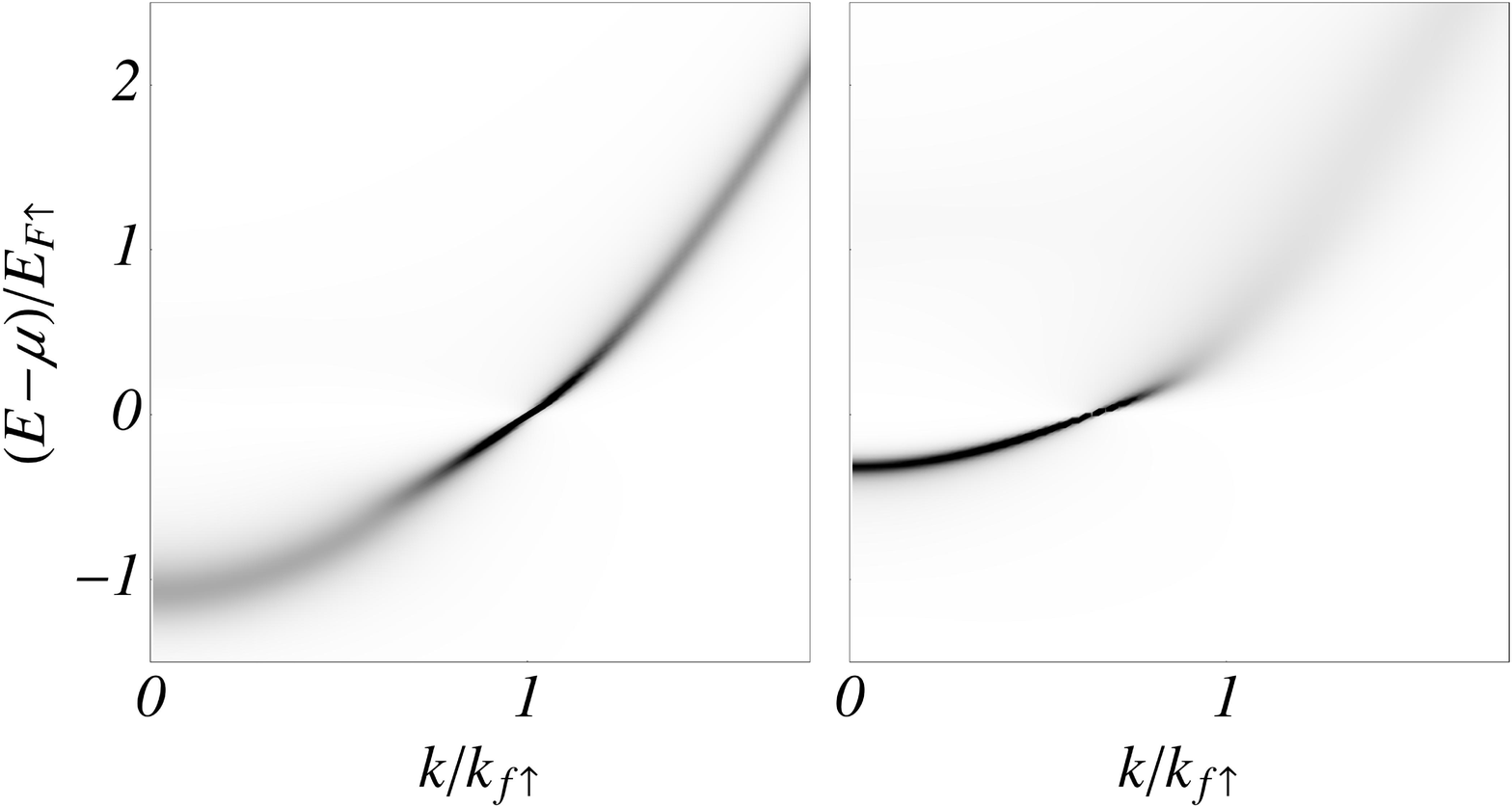}
\caption{Single particle spectral density in the normal state of a T=0 two-component Fermi gas with $1/k_{f\uparrow} a=-0.035$, and $n_\downarrow/n_\uparrow=0.25$.  Left: majority ($\uparrow$), Right: minority ($\downarrow$).}\label{polsd}
\end{figure}
  
\begin{figure}[tb]
 \includegraphics[width=\columnwidth]{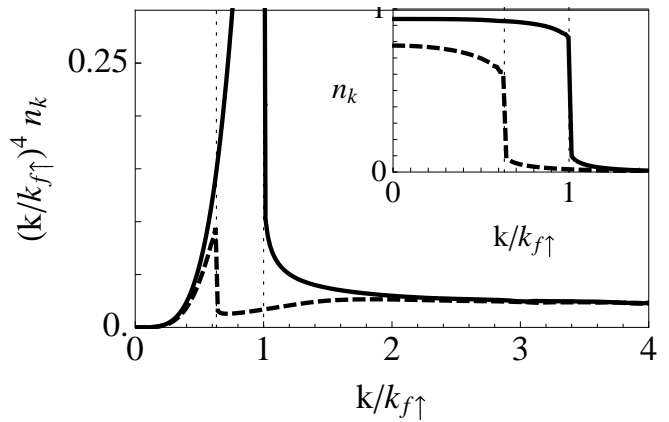}
 \caption{Partially self-consistent T-matrix calculation of  $T=0$ occupation numbers $n_k$ for majority (solid) and minority (dashed) atoms as a function of momentum $k$ scaled by the majority species fermi momentum $k_{f\uparrow}$.  Dotted lines show the location of $k_{f\uparrow}$ and $k_{f\downarrow}$.  Main figure shows occupation numbers scaled by $(k/k_f)^4$ in order to illustrate their asymptotic behavior, while the inset shows the unscaled occupation numbers. Here $1/(k_{f\uparrow} a)=0.35$ and $n_\downarrow/n_\uparrow=0.25$.}\label{occ}
 \end{figure}

\begin{figure}
 \includegraphics[width=\columnwidth]{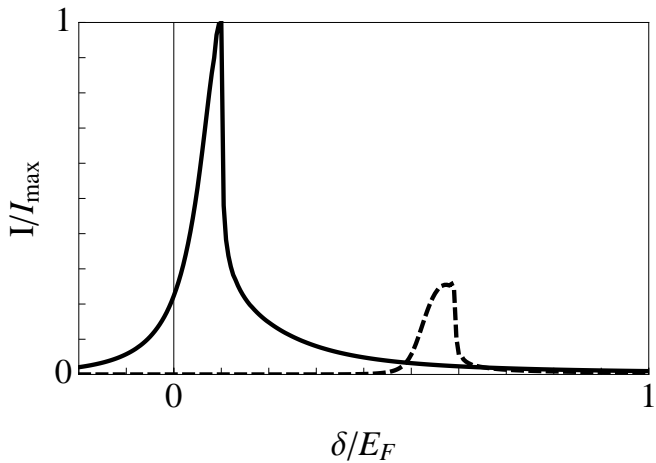}
 \caption{Partially self-consistent T-matrix calculation of  $T=0$ RF spectrum $I$ for majority (solid) and minority (dashed) atoms in the normal state of a polarized Fermi gas with $1/(k_{f\uparrow} a)=0.35$ and $n_\downarrow/n_\uparrow=0.25$.  Spectrum is normalized by the peak value of the majority species intensity $I_{\rm max}$.
 Similar figures appear in \cite{schneider}
 }\label{rf}
 \end{figure}

The use of free propegators in (\ref{tmateqn}) leads to some unphysical results \cite{parish,schneider,liuhu}.  The key problem is that when one considers Pauli blocking in the two-particle collisions, one uses the densities of free particles with chemical potentials $\mu_\uparrow$ and $\mu_\downarrow$.  These densities are much smaller than those of interacting particles at the same chemical potential. To mitigate this problem, 
we follow Schneider et al. \cite{schneider} and  shift $\mu_\sigma$ in (\ref{freeg}) so that $n_\sigma^{(0)}=(\beta\Omega)^{-1}\sum_{km} G^{(0)}_k(\nu_m)$ and   $n_\sigma=(\beta\Omega)^{-1}\sum_{km} G_k(\nu_m)$ are equal.  We view this  as a form of self-consistency \cite{selfconsistentnote}, where we approximate our self-energies as a constant.   Perali et al. use a similar approach, but with a slightly different renormalization procedure \cite{perali}.  While it avoids some unphysical results, the quantitative predictions of the partially self-consistent theory tend to be less accurate than the non-selfconsistent theory.  For example, within our approximations we find $T_c=0.5E_f$.
Except for some additional broadening, the rough qualitative features of Figs.~\ref{dos}-\ref{polsd} are the same with and without self-consistency.  
The spectral tails in Figs.~\ref{occ}-\ref{rf} differ in the two theories.  

 We tabulate the  imaginary part of the self-energy, $\Gamma=2 {\rm Im}(\Sigma)$, via 
 \begin{equation}
 \Gamma_{p\downarrow}(\omega)=\int\frac{d^3k}{(2\pi)^3}\Lambda_{k+p}(\omega+\epsilon_{k\uparrow})\left[
f+g
 \right],
 \end{equation}
 where $\Lambda_{k}(\omega)=2{\rm Im}(T_k(\omega))$, $f=1/(e^{\beta\epsilon_{k\uparrow}}+1)$, and $g=1/(e^{\beta(\omega+\epsilon_{k\uparrow})}-1)$.
We then use the Kramers-Kronig relationship to extract the real part.  
 From the single particle Greens function, $G^{-1}=G_0^{-1}-\Sigma$, we extract the spectral density $A_{k\sigma}(\omega)=2{\rm Im} G_\sigma(k,\omega)$, which is the number of single-particle states which have energy $\omega$ and momentum $k$.   The density of states is $\rho_\sigma(E)=\sum_k A_{k\sigma}(E)$.  Sample density of states are shown in figure~\ref{dos} for temperatures slightly above $T_c$.  As interactions are made stronger one sees a pronounced dip in the density of states near the Fermi surface.  In figure~\ref{sd} the full momentum resolved spectral density is shown.  As interactions grow, the bottom of the band is shifted downward -- an effect analogous to the
  ``Hartree" shift seen at weak coupling.  The higher energy states have a smaller shift.  The break-point, near the Fermi energy, has low spectral weight.  
 If one looks carefully at these figures, especially the third from the left, one sees a faint downward dispersing band.  These states can be interpreted as excitations where one creates a positive energy (ie. non-condensed) ``pair''.  Crudely, one adds a $\uparrow$ particle of momentum $k$ to this band by adding a pair of momentum $p$ (with energy $E_{\rm pair}(p)$) and removing a $\downarrow$ particle of momentum $p-k$.  This changes the system's energy by $E_{\rm pair}(p)-\epsilon_{p-k\downarrow}$.  Assuming that the bulk of the spectral weight comes from the lowest energy pairs, which are at $p=0$ with vanishing energy as $T\to T_c$ (ie. $E_{\rm pair}(0)\approx0$), one should find 
 single particle states near the line $\omega=-\epsilon_{k\downarrow}$.   In analogy to what happens in the superfluid, the hybridization between these ``hole" states and the regular ``particle" states where $\omega=\epsilon_{k\uparrow}$ gives rise to the pseudogap.   On the BEC side of resonance, which we do not discuss here, $E_{\rm pair}(0)<0$, and the entire downward dispersing band lies at negative energies.
 
 This ``hole" branch is quite diffuse in the normal state. Its
  breadth 
  comes from both the fact that one can add pairs with $p\neq0$, and the fact that away from on the BEC side of resonance, the pairs do not have particularly sharp energies.  This latter point is mathematically seen in the structure of the $T$-matrix.  The T-matrix does not have a pole on the real axis, though near $T_c$ it becomes sharply peaked.  In that regime
 our ``cartoon" gives a reasonable picture of the underlying physics.  We must add Schneider et al.'s caveat that one can get downward dispersing spectral weight merely from short range correlations \cite{schneider}, though this argument mainly applies to large wave-vectors.
 On a purely technical note, our partial self-consistency makes these spectra more diffuse than those in \cite{ohashi}.  


 As one polarizes the gas by increasing $\mu_\uparrow$ and decreasing $\mu_\downarrow$, the cartoon argument says that the crossing point moves:  both the $\uparrow$-particle and $\downarrow$-hole branches move to lower energy, so the crossing moves to negative $\omega$ in the  majority ($\uparrow$) spectral density .  For the $\downarrow$-atoms the opposite occurs, and the crossing in the minority spectral density is at positive $\omega$.  This scenario resembles the 
   ``breached pair" states introduced by Liu and Wilczek \cite{breached}, but this is in the normal state.  The physical picture is that there are initially no pairs, but one can create pairs while adding a particle.   As discussed by Fumarola et al \cite{fum}, once you polarize the gas there is a sharp pole in the T-matrix  (but the total weight in this peak is not necessarily large).
 

Our calculation of the density of states is consistent with this ``excited pairs" scenario.  In Fig.~\ref{zerotdos}, one sees that in the $T=0$ polarized normal state the majority species density of states has a slight dip at energies below the Fermi surface, and in the minority spectral density the dip occurs above the Fermi surface.  These dips (particularly the one in the majority density of states) are much more subtle than those in Fig.~\ref{dos}.  In the momentum resolved spectra shown in Fig.~\ref{polsd}, one sees that the dips are caused by the extremely short lifetime of the quasiparticle excitations in those energy ranges.  This behavior is what one would expect from the cartoon of hybridizing the single particle excitations with excited pair states.

 Importantly, the zero temperature polarized normal state is a Fermi liquid.  For example, the excitations near the Fermi surface in Fig.~\ref{polsd} have lifetimes which scale as $\tau\sim(E-\mu)^{-2}$.  The occupation numbers, illustrated in the inset of Fig.~\ref{occ}, show a discontinuity.  As noted by Haussmann \cite{haussmann2} in the early 90's, for large $k$, $n_k\sim k^{-4}$.  This is a consequence of the discontinuity in the slope of the wavefunction when two atoms of unequal spin approach one-another.  More recently, Tan quantified this relationship, relating the coefficient of the power law to the ``contact" \cite{tan}.  The value of this coefficient has been well explored both theoretically and experimentally in the ballanced case \cite{balcontact}. In the imbalanced gas, one expects that for large $k$ the occupation numbers should agree $n_{k\uparrow}=n_{k\downarrow}$.  This large $k$ equality follows from the fact that the short range discontinuity involves the relative wavefunction of an atom of each species.  In the absence of self-consistency, Eq~(\ref{tmateqn}) fail to reproduce this result.  However, in the theory used here (Fig.~\ref{occ}), where we shift the chemical potentials, the tails coincide \cite{schneider}.

 What are the consequences of this structure for experiment?  Since the pseudogap has moved to negative energies in the majority-species spectral density, it can in principle be observed by final-state free  RF-spectroscopy experiments.  Unfortunately, an explicit calculation of the momentum unresolved spectrum,
 %
$I_\sigma(\nu_{\rm rf})=\int d^3k/(2\pi^3)\,A_{k\sigma}(k^2/2m-\mu_\sigma-\nu_{\rm rf}) f(k^2/2m-\mu_\sigma-\nu_{\rm rf})$
does not show any particularly distinct feature which can be unambiguously attributed to the pseudogap (Fig.~\ref{rf}).  Momentum resolved spectra, which give information analogous to photo-emission \cite{latestjin},  directly measure $A_{k\sigma}(\omega)$. As shown in Fig.~\ref{sd}, the finite energy pseudogap is apparent in this function as a strong energy dependence in the quasiparticle lifetime.

 As seen in Fig.~\ref{rf}, at large $\delta$ the minority and majority rf spectra coincide.  This is a consequence of the fact that the tails of the spectrum correspond to short range physics, and is a measure of Tan's contact (as derived by several authors \cite{schneider,pierinatphys,braatenkang}, it has the form $I(\nu)=C \nu^{-3/2}/4\pi^2\sqrt{2 m}$).  As with the tails of the occupation numbers (Fig.~\ref{occ}), the tails of the majority and minority spectra fail to coincide if one does not adjust the chemical potentials \cite{schneider}.
Note that as one moves to the BEC side of the resonance, one experimentally sees that the majority species RF spectrum becomes bimodal \cite{zw}, but near unitarity, where our calculations were performed, the experimental spectrum is unimodal.
%

 \section*{Acknowledgements}
 
 I would like to thank Kathy Levin for extensive discussions, as well as input from Sourish Basu, Stefan Baur, Kaden Hazzard, and Stefan Natu. This material is based upon work supported by the National Science Foundation through grant No PHY-0758104.

 \end{document}